\documentclass[prl,aps,showpacs,twocolumn]{revtex4}
\usepackage{amssymb}

\usepackage{graphicx}
\usepackage{bm}

\begin{document}

\title{Semiclassical Dynamics of Electrons in Magnetic Bloch Bands: a Hamiltonian Approach}
\author{Pierre Gosselin$^{1}$, Fehrat M\'enas$^{2}$, Alain B\'{e}rard$^{3}$, Herv\'{e} Mohrbach$^{3}$}
\address{$^1$Institut Fourier, UMR
5582 CNRS-UJF, UFR de Math\'ematiques, Universit\'e Grenoble I, BP74, 38402 Saint Martin
d'H\`eres, Cedex, France \\
$^2$ Laboratoire de Physique et de Chimie Quantique, Facult\'e des Sciences, Universit\'e Mouloud Mammeri,  BP 17 Tizi Ouzou, Algerie\\
$^3$ Laboratoire de Physique Mol\'eculaire et des Collisions,
ICPMB-FR CNRS 2843, \\ Universit\'e Paul Verlaine-Metz,  1
boulevard Arago, 57078 Metz Cedex 3, France}


\begin{abstract}
By formally diagonalizing with accuracy $\hbar$ the
Hamiltonian of electrons in a crystal subject to electromagnetic
perturbations, we resolve the debate on the Hamiltonian nature of
semiclassical equations of motion with Berry-phase corrections,
and therefore confirm the validity of the Liouville theorem. 
We show that both the position and momentum operators acquire a Berry-phase dependence, leading to a non-canonical Hamiltonian
dynamics. The
equations of motion turn out to be identical to the ones previously
derived in the context of electron wave-packets dynamics.

\end{abstract}

\maketitle

The notion of Berry phase has found many applications in several
branches of quantum physics, such as atomic and molecular physics,
optic and gauge theories, and more recently, in spintronics, to cite
just a few. Most studies focused on the geometric phase a
wave function acquires when a quantum mechanical system has an
adiabatic evolution. It is only recently that a possible influence
of the Berry phase on semiclassical dynamics of several physical
systems has been investigated. It was then shown that Berry phases
modify semiclassical dynamics of spinning particles in electric
\cite{ALAIN} and magnetic fields \cite{BLIOKH1}, as well as in
semiconductors \cite{MURAKAMI1}. In the above cited examples, a
noncommutative geometry, originating from the presence of a Berry
phase, which turns out to be a spin-orbit coupling, underlies the
semiclassical dynamics. Also, spin-orbit contributions to the
propagation of light have been the focus of several other works
\cite{ALAIN,BLIOKH2, MURAKAMI2}, and have led to a generalization of
geometric optics called geometric spinoptics \cite{HORVATHY1}.

Semiclassical methods in solid-state physics have also played an
important role in studying the dynamics of electrons to account
for the various properties of metals, semiconductors, and
insulators \cite{MERMIN}. In a series of papers \cite{NIU1,NIU3}
(see also \cite{SHINDOU}), the following new set of semiclassical
equations with a Berry-phase correction was proposed to account
for the semiclassical dynamic of electrons in magnetic Bloch bands
(in the usual one-band approximation)
\begin{eqnarray}
\dot{\mathbf{r}}&=&\partial \mathcal{E}(\mathbf{k})/\hbar
\partial
\mathbf{k}- \dot{\mathbf{k}} \times\Theta(\mathbf{k}) \nonumber \\
\hbar\dot{\mathbf{k}}&=&-e\mathbf{E}-e\dot{\mathbf{r}}\times
\mathbf{B(r)} \label{niuequations}
\end{eqnarray}
where $\mathbf{E}$ and $\mathbf{B}$ are the electric and magnetic
fields respectively and
$\mathcal{E}(\mathbf{k})=\mathcal{E}_0(\mathbf{k})-\mathbf{m(k).B}$
is the energy of the $n$th band with a correction due to the
orbital magnetic moment \cite{NIU3}. The correction term to the
velocity $- \dot{\mathbf{k}} \times \Theta$ with
$\Theta(\mathbf{k})$ the Berry curvature of electronic Bloch state
in the $n$th band is known as the anomalous velocity predicted to
give rise to a spontaneous Hall conductivity in ferromagnets
\cite{LUTTINGER}. For crystals with broken time-reversal symmetry
or spatial inversion symmetry, the Berry curvature is nonzero
\cite{NIU3}. Eqs.\ref{niuequations} were derived by considering a
wave packet in a band and using a time-dependent variational
principle in a Lagrangian formulation. The derivation of a
semiclassical Hamiltonian was shown to lead to difficulties in the
presence of Berry-phase terms \cite{NIU3}. The apparent
non-Hamiltonian character of Eqs.\ref{niuequations} led the authors
of \cite{NIU2} to conclude that the naive phase space volume is
not conserved in the presence of a Berry phase, thus violating
Liouville's theorem. To remedy this situation these authors
introduced a modified density of state in the phase space
$D(\mathbf{r},\mathbf{k})=(2\pi)^{-d}(1+e\mathbf{B}.\Theta/\hbar)$
such that $D(\mathbf{r},\mathbf{k})d\mathbf{r} d\mathbf{k}$
remains constant in time.

This point of view was immediately criticized by several authors.
In particular, by relating the semiclassical dynamics of Bloch
electron to exotic Galilean dynamics introduced independently in
the context of noncommutative quantum mechanics \cite{HORVATHY2},
C. Duval et al. \cite{HORVATHY3} found that Eqs.\ref{niuequations}
are indeed Hamiltonian in a standard sense, restoring the validity
of Liouville's theorem when the correct symplectic volume form is
used. This approach, relying on a symplectic structure on a
classical Hamiltonian formulation, though very elegant, does not
stem from the quantum Hamiltonian for electrons in a solid and is
consequently not widely known in the solid-state physicists
community. Additionally, the role of the Berry phase is hidden in
this approach. In a different but related work \cite{BLIOKH3}, the
Hamiltonian nature of semiclassical equations of motion of Dirac
electrons in electromagnetic field with Berry-phase corrections
(in this case it corresponds to a spin-orbit coupling) was
established.

This letter presents an alternative approach for the
derivation of the equations of motion of an electron in magnetic
Bloch bands, based on a direct semiclassical diagonalization of
the full quantum Hamiltonian. We show that both dynamical
variables ($\mathbf{r}, \mathbf{k}$) in Eqs. \ref{niuequations} are
not the usual Galilean operators, but new covariant operators defined in a particular $n$th Bloch band, and including Berry gauge potentials. These potentials induce non-canonical commutation relations between the covariant variables. In our context, the equations of motion are given by the tandard
dynamical laws $\hbar d\mathbf{r}/dt=i\left[ \mathbf{r},H\right] $
and $d\mathbf{k}/dt=i[\mathbf{k} ,H] $ leading to
Eqs.\ref{niuequations} in a semiclassical approximation. Our
approach thus reveals the Hamiltonian nature of
Eqs.\ref{niuequations} and confirms the importance of the Berry
phase on the semiclassical dynamics of Bloch electrons. The origin
of the density of state $D(\mathbf{r},\mathbf{k})$ is then
obvious; it is simply equal to the Jacobian of the transformation
between the canonical variables $(\mathbf{R},\mathbf{K})$ and the
covariant ones $(\mathbf{r},\mathbf{k})$, as already found in the
context of the Dirac equation in \cite{BLIOKH3}.

It should be noted that focusing on a Hamiltonian formalism for
electrons in solids in order to account for the anomalous velocity
was first initiated by Adams and Blount \cite{BLOUNT}, who showed
that this term arises from the noncommutability between the
components of the intraband position operator, which acquires a
Berry-phase contribution. But their approach does not lead to the
correct Eqs.\ref{niuequations} for electrons in magnetic
Bloch waves, as they missed the Berry phase dependence of the
intraband momentum operator. A similar Hamiltonian approach has
also been realized for arbitrary spinning (massive and massless)
particles in an electric field \cite{ALAIN} and extended to the
case of Dirac electron in an arbitrary electromagnetic field
\cite{BLIOKH1,BLIOKH3}. The common feature of these Hamiltonian
formulations is that a noncommutative geometry underlies the
algebraic structure of both coordinates and momenta. Actually, a
Berry-phase contribution to the coordinate operators stems from
the representation where the kinetic energy is diagonal
(Foldy-Wouthuysen or Bloch representation). The components of the
coordinate become noncommutative when interband transitions are
neglected (adiabatic motion).

Consider an electron in an crystal lattice perturbated by the
presence of an external electromagnetic field. As is usual, we
express the total magnetic field as the sum of a constant field
$\mathbf{B}$ and small nonuniform part $\delta \mathbf{B}(\mathbf{R})$. The
Hamiltonian can be written $H=H_0 -e\phi(\mathbf{R})$, with $H_0$
the magnetic contribution ($\phi$ being the electric potential)
which reads
\begin{equation}
H_0=\left( \frac{\mathbf{P}}{2m}+e\mathbf{A}(\mathbf{R})+ e\delta
\mathbf{A}(\mathbf{R})\right) ^{2}+V(\mathbf{R})
\label{Hmagnetic}
\end{equation}
where $\mathbf{A}(\mathbf{R})$ and $\delta \mathbf{A}(\mathbf{R})$
are the vectors potential of the homogeneous and inhomogeneous
magnetic field, respectively, and $V(\mathbf{R})$ the periodic
potential. The large constant part $\mathbf{B}$ is chosen such
that the magnetic flux through a unit cell is a rational fraction of
the flux quantum $h/e$. The
advantage of such a decomposition is that for $\delta
\mathbf{A}(\mathbf{R})=0$ the magnetic translation operators
$\mathbf{T}(\mathbf{R}_{i})=\exp(i\mathbf{K}.\mathbf{R}_{i})$, with $\mathbf K$
the generator of translation, are commuting quantities
allowing to exactly diagonalize the Hamiltonian and to treat $\mathbf{\delta
}A(\mathbf{R})$ as a small perturbation. The state space of the Bloch
electron is spanned by the basis vector
$|n,\mathbf{k}>=|\mathbf{k}>\otimes|n> $ with $n$ corresponding to a band indice.
In this representation $\mathbf K|n,\mathbf{k}>=\mathbf k|n,\mathbf{k}>$ and
the position operator is $\mathbf R=i\partial /\partial \mathbf K$, which
implies the canonical commutation relation $[R_i,K_j]=i\delta_{ij}$.

We first perform the diagonalization of the Hamiltonian in
Eq.\ref{Hmagnetic} for $\delta \mathbf{A}=0$ by an unitary matrix
$U(\mathbf{K})$ (whose precise expression is not necessary for the
derivation of the equations of motion) such that
$UHU^+=\mathcal{E}(\mathbf{K})-e\phi(U\mathbf{R}U^+)$,
where $\mathcal{E}(\mathbf{K})$ is the diagonal energy matrix made
of elements $\mathcal{E}_n(\mathbf{K})$ with $n$ the band indice.
Whereas the quasi-momentum is invariant through the action of $U$,
e.g., $\mathbf{k}=U\mathbf{K}U^+=\mathbf{K}$, the position operator
becomes:
\begin{equation}
\mathbf{r} = U\mathbf{R}U^{+}= \mathbf{R}+i U
\partial _{\mathbf{k}}U^{+}  \label{r}
\end{equation}
in the new representation owing to the fact that $[R_i, K_j]=i
\delta_{i,j}$. In the adiabatic or one-band approximation, which
consists of neglecting interband transitions, one has to project
the position coordinate (the momenta operator is diagonal and so
invariant by construction) on a certain band such that the $n$th
intraband position operator $\mathbf{r}_n =
\mathcal{P}_n(\mathbf{r})$ reads $\mathbf{r}_n = \mathbf{R}+
\mathcal{A}_{n}$. The quantity $\mathcal{A}_{n}=i\mathcal{P}_n(U
\partial _{\mathbf{k}}U^{+})$ is a Berry connection, as it can
be readily shown that its matrix elements are given by
$\mathcal{A}_{n}(\mathbf{k})=i <u_n(\mathbf{k}) |\partial
_{\mathbf{k}}|u_n(\mathbf{k})>$, where we used $U^+(\mathbf{k})|n>=|u_n(\mathbf{k})>$ with
$|u_n(\mathbf{k})>$ the periodic
part of the magnetic Bloch waves. The price to pay when considering
the one-band approximation is that the algebra of the coordinates
becomes noncommutative (as we consider only one band, we drop the
index $n$)
\begin{equation}
\left[ r^{i},r^{j}\right] =i\Theta^{ij}(\mathbf{k}) \label{RR}
\end{equation}
with
$\Theta^{ij}(\mathbf{k})=\partial^i\mathcal{A}^{j}(\mathbf{k})-
\partial^j\mathcal{A}^{i}(\mathbf{k})$ the Berry curvature.
Observe that the replacement of $\mathbf k$ by $\mathbf{p}/\hbar$
shows that $\Theta^{ij}(\mathbf{p})$ is actually of order
$\hbar^2$. In the one-band approximation the full Hamiltonian,
including the electric potential, is now given by
\begin{equation}
\mathcal{P}_n(UHU^+)=\mathcal{E}(\mathbf{k})-e\phi(\mathbf{r}).
\end{equation}
Due to the Berry connection in the definition of the position
operator, the equations of motion should be changed. But to compute
commutators like $\left[ r^{k},\phi(\mathbf{r}) \right] $, one
resorts to the semiclassical approximation $\left[ r^{k},
\phi(\mathbf{r}) \right] =i\partial _{l}\phi(\mathbf{r})\Theta
^{kl}+O(\hbar )$ leading to the following semiclassical equations
of motion
\begin{equation}
\dot{\mathbf{r}}=\partial\mathcal{E}(\mathbf{k})/\hbar\partial\mathbf{k}
- \dot{\mathbf{k}} \times\Theta(\mathbf{k}), \text{ }
\hbar\dot{\mathbf{k}}=-e\mathbf{E}
\end{equation}
where $\mathbf{E}$ is the external electric field. Whereas the
momentum equation of motion is the usual one, the velocity
operator acquires an anomalous contribution due to presence of the
Berry curvature. Notice that the contribution of the magnetic
field stems only from the presence of the Berry curvature through
the band structure. This equation was first derived by Adams and
Blount \cite{BLOUNT} using a similar approach, and later rederived
by Niu and coworkers \cite{NIU1,NIU3} by looking at the dynamics
of wave packets from a Lagrangian formalism. In the following, we
will extend our approach to carry out a semiclassical
diagonalization of the full electromagnetic Hamiltonian (with
$\delta \mathbf{A}(\mathbf{R})\neq 0$). Contrary to the work of
\cite{BLOUNT}, we show that the momentum also acquires a
Berry-phase contribution leading to different semiclassical
equations of motion. These last ones turn out to be those derived
first in \cite{NIU1,NIU3} (also Duval et al. \cite{HORVATHY3} in
another context). Our rigorous approach has the merit to show
without ambiguities that the equations of motion are indeed
Hamiltonian in the standard sense.

The diagonalization of the Hamiltonian in the presence of an
arbitrary magnetic field is now the focus of the rest of the
paper. Consider first the Hamiltonian Eq.\ref{Hmagnetic} in the
absence of an electric field and set $ \tilde{\mathbf{K}}
=\mathbf{K}+e\delta \mathbf{A}(\mathbf{R})/\hbar$. As the flux
$\mathbf{\delta B}$ on a plaquette is not a rational multiple
of the flux quantum, we
cannot diagonalize simultaneously its components $\tilde{K}_{i}$
since they do not commute anymore. Actually
\begin{equation}
\hbar [ \tilde{K}^{i},\tilde{K}^{j}] =-ie\varepsilon ^{ijk}\delta
B_{k}( \mathbf{R})
\end{equation}
As a consequence of this non-commutativity, we just aim at
quasi-diagonalizing our Hamiltonian at the semiclassical order
(with accuracy $\hbar$). To perform this approximate
diagonalization $\tilde{U}H\tilde{U}^{+}$ with accuracy $\hbar$ we
first consider the limiting case of a constant potential  $\delta
\mathbf{A}(\mathbf{R})=\delta \mathbf{A}_0$ (this is obviously a
formal consideration). Clearly, the Hamiltonian in Eq.\ref{Hmagnetic}
is diagonalized by the matrix $U\left( \mathbf{\delta A}\right)
=U(\mathbf{K}+e\mathbf{\delta A}/\hbar)$, as we have just
shifted the momentum $\mathbf{K}$. To diagonalize Eq.
\ref{Hmagnetic} in the general case, we now consider a unitary
matrix $\tilde{U}(\mathbf{K}+e\delta
\mathbf{A}(\mathbf{R})/\hbar)$ which has the same series expansion
as $U\left( \delta \mathbf{A}\left( \mathbf{R}\right) \right) $
when $\mathbf{R}$ is considered as a parameter commuting with
$\mathbf{K}$. Of course, this matrix is not unique, due to the
noncommutativity of $\mathbf{K}$ and $\mathbf{R}$, but it can be
shown that the various choices lead to the same projected
Hamiltonian. Note that in the sequel, a small $\delta \mathbf{A}$
perturbation, which preserves the band structure determined
previously is assumed, i.e. $<n|\delta \mathbf{A}|m>=0$ for $m\neq
n$. Before implementing effectively the canonical transformation
on the Hamiltonian, it appears more convenient to implement
first the canonical transformation on the dynamical operators.
Therefore, in the new representation the position operator is
again given by $\mathbf{r} = \mathbf{R}+i \tilde{U}
\partial_{\mathbf{\tilde{K}}}\tilde{U}^{+}$.
As before, the projection on a band defined the $n$th intraband
position operator $\mathbf{r}_n =\mathbf{R}+
\mathcal{A}_n(\tilde{\mathbf{K}})$, with
$\mathcal{A}_n(\tilde{\mathbf{K}})=\mathcal{P}_n(
\tilde{U}\partial_{\tilde{\mathbf{K}}}\tilde{U}^{+})$ a new Berry
connection.

The pseudo-momentum $\mathbf{\tilde{K}}$ is no more invariant
as we obtain
\begin{eqnarray}
\tilde{\mathbf{k}} &=&\tilde{U}\tilde{\mathbf{K}}\tilde{U}^{+}=
\tilde{\mathbf{K}}+\tilde{U}\partial
_{\tilde{K}^{j}}\tilde{U}^{+}\left[
\tilde{\mathbf{K}},\tilde{K}^{j}\right]  \nonumber \\
&=&\tilde{\mathbf{K}}-ie\tilde{U}\partial
_{\tilde{\mathbf{K}}}\tilde{U}^{+}\times \delta
\mathbf{B}(\mathbf{R})/\hbar \label{momenta}
\end{eqnarray}
The $n$th intraband momentum operator
$\tilde{\mathbf{k}}_n=\mathcal{P}_n(\tilde{\mathbf{k}})$ is then
\begin{equation}
\hbar\tilde{\mathbf{k}}_{n}=\hbar\tilde{\mathbf{K}}-e\mathcal{A}_n(\tilde{\mathbf{K}})\times\delta
\mathbf{B}(\mathbf{R})
\end{equation}
which at the order $\hbar$ can also be written
\begin{equation}
\hbar \tilde{\mathbf{k}}_{n}\simeq \hbar\tilde{\mathbf{K}}-e
\mathcal{A}(\tilde{\mathbf{k}}_n)\times \delta
\mathbf{B}(\mathbf{r}_n) +O(\hbar^2)
\end{equation}
This new contribution to the momentum has been overlooked before
in the work of Adams and Blount \cite{BLOUNT} but is crucial for
the correct determination of the semiclassical equations of motion
of an electron in a magnetic Bloch band.

The commutation relations between the components of the intraband
momenta are therefore given by (at leading order)
\begin{equation}
\hbar \left[ \tilde{k}_n^i,\tilde{k}_n^{j}\right] =-ie\varepsilon
^{ijk}\delta B_{k}(\mathbf{r}_n)+ ie^2\varepsilon^{ipk}\delta B_{k}\varepsilon^{jql}\delta B_{l}\Theta^{pq}/\hbar
\label{KK}
\end{equation}
with $\Theta^{ij}(\tilde{\mathbf{k}}_n)=\partial^i\mathcal{A}^{j}
(\tilde{\mathbf{k}}_n)-\partial^j\mathcal{A}^{i}(\tilde{\mathbf{k}}_n)$
the Berry curvature. The commutation relation between position and
momentum can be computed leading to
\begin{equation}
\left[r_n^{i}, \hbar \tilde{k}_n^{j}\right] = i\hbar\delta
^{ij}+ie\varepsilon ^{jlk} \delta
B_{k}(\mathbf{r}_n)\Theta^{il}(\tilde{\mathbf{k}}_n) \label{RKK}
\end{equation}
The third useful commutator is as in Eq.\ref{RR} given by
\begin{equation}
\left[ r_n^{i},r_n^{j}\right] = i\Theta(\tilde{\mathbf{k}}_n)
^{ij}  \label{RRR}
\end{equation}
at leading order. The set of nontrivial commutations relations
given by Eqs.\ref{KK}, \ref{RKK}, \ref{RRR} is the same as the one
deduced in \cite{BLIOKH3} in the context of the Dirac electron
using an approximate explicit Foldy Wouthuysen transformation.
This shows that a common structure underlies the
quasi-diagonalization of general quantum Hamiltonians in the
presence of electromagnetic fields \cite{PIERRE}. In the present
case, the approximate diagonalization $\tilde{U}H\tilde{U}^{+}$ is
performed by formally expanding $\tilde{U}$ and $H$ in a series of
$\mathbf{K}$ and $\delta \mathbf{A}(\mathbf{R}) $. The
recombination of the series includes corrections of order $\hbar$
due to the noncommutativity of $\mathbf{K}$ and $\mathbf{R}$.
In doing so, we arrive at the following expression
\begin{eqnarray}
\tilde{U}H\tilde{U}^{+}&=&\mathcal{E}\left( \tilde{\mathbf{k}}\right)
-\frac{ie}{4\hbar }
\left[ \mathcal{E} (\tilde{\mathbf{K}}),\mathcal{A}_i(\tilde{\mathbf{K}})\right] \varepsilon ^{ijk}
\delta B^{k}(\mathbf{R})\mathcal{A}_j(\tilde{\mathbf{K}}) \nonumber\\
&&-
\frac{ie}{4\hbar}\mathcal{A}_j(\tilde{\mathbf{K}})\left[ \mathcal{E} (\tilde{\mathbf{K}}),\mathcal{A}_i(\tilde{\mathbf{K}})
\right] \varepsilon ^{ijk}\delta B^{k}(\mathbf{R}) \nonumber
\end{eqnarray}
which after projection on the $n$th band can be written:
\begin{equation}
\mathcal{P}_n(\tilde{U}H\tilde{U}^{+})=\mathcal{E}_n \left(
\tilde{\mathbf{k}}_n\right) -
\mathcal{M}(\tilde{\mathbf{K}}).\delta \mathbf{B}(\mathbf{r}_n)
+O(\hbar^2)
\end{equation}
with $\mathcal{M}(\tilde{\mathbf{K}})=\mathcal{P}_n(
\frac{ie}{2\hbar}\left [\mathcal{E}(\tilde{\mathbf{K}}),
\mathcal{A}(\tilde{\mathbf{K}}) \right]
\times\mathcal{A}(\tilde{\mathbf{K}}))$ the magnetization. This
term can also be written under the usual form in the
$(\mathbf{K},n)$ representation \cite{LANDAU}:
\begin{eqnarray}
\mathcal{M}^i_{nn}=\frac{ie}{2\hbar}\varepsilon^{ijk} \sum_{n'\neq n} (\mathcal{E}_n-\mathcal{E}_{n'})(\mathcal{A}_j)_{nn'}(\mathcal{A}_k)_{n'n}
\end{eqnarray}
We mention that this magnetization (the orbital magnetic moment of Bloch electrons), has been obtained previously 
in the context of electron wave packets dynamics \cite{NIU1,NIU3}. 

\smallskip

Notice that because a semiclassical computation was considered here, we kept only terms of order $\hbar$. As $\delta \mathbf{A}$ is small, we
chose to neglect terms of order $\hbar\delta \mathbf{A}^2$. But,
as we do not consider a perturbation expansion, our method keeps
all contributions of order $\delta \mathbf{A}^n$. In a
perturbation expansion, instead of evaluating
$\tilde{U}H\tilde{U}^{+}$, one would compute
$U(\mathbf{K})HU(\mathbf{K})^+=\mathcal{E}(\mathbf{K})+U\delta
HU^{+}$ (and neglect all terms of order higher than $\delta
\mathbf{A}$ ). In this representation the position operator is
still given by Eq.\ref{r} but $\mathbf K$ is invariant. But doing
so would lead us to neglect contributions of order $\hbar$ that
are fundamental for the correct determination of the equations of
motion. A perturbation expansion is then not allowed here.
\smallskip

The commutation relations Eqs.\ref{KK}, \ref{RKK}, \ref{RRR},
together with the semiclassical Hamiltonian of the Bloch electron
in the full electromagnetic field
$E_n(\tilde{\mathbf{k}}_n)-\phi(\mathbf{r}_n)$ with
$E_n(\tilde{\mathbf{k}}_n)=
\mathcal{E}_n(\tilde{\mathbf{k}}_n)-\mathcal{M}(\tilde{\mathbf{k}}_n).\delta
\mathbf{B}(\mathbf{r}_n)$, allow us to deduce the semiclassical
equations of motion. Dropping now the index $n$ we have:
\begin{eqnarray}
\dot{\mathbf{r}}&=&\partial E(\tilde{\mathbf{k}})/\hbar
\partial \tilde{\mathbf{k}}- \dot{\tilde{\mathbf{k}}}
\times \Theta(\tilde{\mathbf{k}}) \nonumber \\
\hbar\dot{\tilde{\mathbf{k}}}&=&-e\mathbf{E}-e\dot{\mathbf{r}}\times
\delta\mathbf{B}(\mathbf{r})-\mathcal{M}.\partial
\delta\mathbf{B}/\partial\mathbf{r} \label{EQM}
\end{eqnarray}
These equations differ from the ones derived in \cite{BLOUNT}, but
are exactly the same equations of motion as in \cite{NIU1,NIU3} apart from
the magnetization contribution (which should
also be present in \cite{NIU3}). It is also clear that the
correct volume form in the phase space ($\mathbf{r},
\mathbf{\tilde{k}})$ has to include the Jacobian $D(\mathbf{r},
\mathbf{\tilde{k}})=(1+e\delta
\mathbf{B}.\Theta/\hbar)$ of the transformation from ($\mathbf{R},
\mathbf{\tilde{K}}$) to ($\mathbf{r}, \mathbf{\tilde{k}})$. This
Jacobian is precisely the density of state introduced in
\cite{NIU2}, in order to ensure the validity of the
Liouville theorem. As a consequence, and by comparing
Eqs.\ref{niuequations} and \ref{EQM} we can
conclude that the operators 
($\mathbf{r}, \mathbf{\tilde{k}}$) correspond to the dynamical variables in Eqs. \ref{niuequations}, denoted $\mathbf{x_{c}}$ and $\mathbf{q_{c}}$ in Ref. \cite{NIU3}. The variable $\mathbf{x_{c}}$ is the
wave-packet center, and $\mathbf{q_{c}}$ the mean wave vector.
The equations of motion for $\mathbf{x_{c}}$ and $\mathbf{q_{c}}$ were
obtained, using a time-dependent variational principle in a
Lagrangian formulation \cite{NIU3} . It was then found that the derivation of a
semiclassical Hamiltonian presents some difficulties in the presence of
Berry-phase terms. Actually, as explained in \cite{NIU3}, this derivation
requires the knowledge of the commutation relations between $\mathbf{x_{c}}$
and $\mathbf{q_{c}}$ (a re-quantization procedure), but these relations
cannot be found from the Lagrangian formulation. One of the advantages of
our approach is to show that these commutation relations are in fact a
direct consequence of the semiclassical diagonalization of the quantum
Hamiltonian.

In summary, our semiclassical diagonalization of the
electromagnetic Bloch Hamiltonian leads to a well defined
semiclassical Hamiltonian with Berry-phase corrections. The
resulting semiclassical equations turn out to be the ones obtained
previously from a semiclassical Lagrangian formalism \cite{NIU3}.
When the correct dynamical variables are used the Liouville
theorem is restored. Moreover, the present approach also confirms the result
of Duval et al. \cite{HORVATHY1} and Bliokh \cite{BLIOKH3} about
the Hamiltonian nature of these semiclassical equations of motion
with Berry-phase corrections, which is a hotly debated subject.  

We would like to thank Aileen Lotz for a critical reading
of the manuscript, and one referee whose pertinent questions
allowed us to improve the present article.


\begin{thebibliography}{99}



\bibitem{ALAIN}  A.\ B\'{e}rard, H.\ Mohrbach, Phys. Rev. D \textbf{69}
(2004) 127701; Phys. lett. A \textbf{352} (2006) 190.

\bibitem{BLIOKH1}  K. Y. Bliokh, Eur. Lett. \textbf{72} (2005) 7.

\bibitem{MURAKAMI1}  S. Murakami, N. Nagaosa, S. C. Zhang, Science \textbf{301}
(2003) 1348.

\bibitem{BLIOKH2}  K. Y. Bliokh, Y. P. Bliokh, Phys. Lett A \textbf{333} (2004), 181;
Phys. Rev. E \textbf{70} (2004) 026605.

\bibitem{MURAKAMI2}M. Onoda, S. Murakami, N. Nagasoa, Phys. Rev. Lett. \textbf{93}
(2004) 083901.

\bibitem{HORVATHY1} C. Duval, Z. Horv\'ath, P. A. Horv\'athy,  Phys. Rev. D \textbf{74} (2006) 021701; and math-ph/0509031.

\bibitem{MERMIN} N. W. Ashcroft and N. D. Mermin, Solid State
Physics (Saunders, Philadelphia, 1976).

\bibitem{NIU1} M. C. Chang, Q. Niu, Phys. Rev. Lett \textbf{75} (1995)
1348; Phys. Rev. B \textbf{53} (1996) 7010;

\bibitem{NIU3} G. Sundaram, Q. Niu, Phys. Rev. B \textbf{59} (1999) 14915.

\bibitem{SHINDOU} R. Shindou, K. I. Imura, Nucl. Phys. B,
\textbf{720} (2005) 399.

\bibitem{LUTTINGER} R. Karplus and J. M. Luttinger, Phys. Rev. \textbf{95}
(1954) 1154; W. Kohn and J. M. Luttinger, Phys. Rev. \textbf{108}
(1957) 590.

\bibitem{NIU2} D. Xiao, J. Shi and Q. Niu, Phys. Rev. Lett \textbf{95} (2005) 137204.

\bibitem{HORVATHY2} P. A. Horv\'athy, L. Martina, P. Stichel,
Phys. Lett. B \textbf{615} (2005) 87.

\bibitem{HORVATHY3} C. Duval, et al., Phys.Rev.Lett. \textbf{96} (2006) 099701; Mod. Phys. Lett. B \textbf{20}, 373 (2006)

\bibitem{BLIOKH3}  K. Y. Bliokh, Phys. Lett. A \textbf{351} (2006) 123.

\bibitem{BLOUNT} E. N. Adams and E. I. Blount, J. Phys. Chem. Solids \textbf{10} (1959) 286.

\bibitem{PIERRE} P. Gosselin, A. B\'erard, H. Mohrbach, hep-th/0603192.

\bibitem{LANDAU} E. M. Lifshitz and L. P. Pitaevskii, Statistical Physics, vol 9, Pergamon Press, 1981.

\end{thebibliography}
\end{document}